\documentstyle[aps,multicol,prl]{revtex}

\draft

\begin{document}
\title{Transport properties of Bi$_2$Sr$_2$CaCu$_2$O$_8$ 
crystals with and without
surface barriers}
\author{D. T. Fuchs$^{1}$, R. A. Doyle$^{2}$, E. Zeldov$^{1}$, 
S. F. W. R. Rycroft$^{2}$, 
T. Tamegai$^{3}$, S. Ooi$^{3}$, M. L. Rappaport$^{4}$, and Y.
Myasoedov$^{1}$}
\address{$^{1}$Department of Condensed Matter Physics,
The Weizmann Institute of Science, Rehovot 76100, Israel}
\address{$^{2}$Interdisciplinary Research Center in
Superconductivity, University of Cambridge, Cambridge CB3 OHE, England}
\address{$^{3}$Department of Applied Physics, The University of
Tokyo, Hongo, Bunkyo-ku, Tokyo, 113, Japan}
\address{$^{4}$Physics Services,
The Weizmann Institute of Science, Rehovot 76100, Israel}
\date{\today}
\maketitle

\begin{abstract}
Bi$_{2}$Sr$_{2}$CaCu$_{2}$O$_{8}$ 
crystals with electrical 
contacts positioned far from the edges are studied by transport 
measurements, then cut into narrow strip geometry, and 
remeasured. Instead of showing larger resistance, the strips 
display a dramatic drop in the resistance, enhanced 
activation energy, and nonlinear behavior due to strong 
surface barriers. The surface barriers also dominate the resistive 
drop at the first-order phase transition. Because the surface 
barriers are avoided in large crystals, we are able to probe the 
solid phase and find good agreement with the recent predictions of 
Bragg glass theory.
\end{abstract}


\begin{multicols}{2}


The resistive behavior of high-temperature superconductors (HTSC) is of
central importance for both fundamental and applied research. In particular,
the transport properties of single crystals of HTSC have been intensively
studied to derive valuable information on critical currents, vortex pinning,
dissipation mechanisms, I-V characteristics, phase transitions, etc. \cite
{review}. These studies are mainly aimed at resolving various aspects of
vortex dynamics in the bulk of the superconductors. The samples are usually
chosen or cut to be in a long and narrow strip geometry for four probe
transport measurements. Longer samples improve the measurement sensitivity,
whereas narrower width is usually assumed to ensure a uniform current flow
and the ability to achieve higher current densities. In this Letter we
demonstrate that this classical geometry turns out to be undesirable for
determining the bulk resistive properties, and may result in apparent
resistivities which are orders of magnitude lower than the true bulk
resistivity.

The current distribution across Bi$_{2}$Sr$_{2}$CaCu$_{2}$O$_{8}$ (BSCCO)
crystals has been recently analyzed by measuring the self-induced magnetic
field of the transport current \cite{sf,pd}. It was found that over a wide
range of temperatures and fields the transport current flows predominantly
at the sample edges\cite{sf,pd}, where vortices enter and exit the crystal,
due to the presence of surface barriers 
(SB)\cite{bl,Clem,burlachkov,GBClem}. 
This finding implies that transport measurements may result in a highly
erroneous evaluation of the bulk resistivity. The self-field technique,
however, does not measure the value of the resistivity and cannot provide
quantitative information on the extent of the error. For that purpose the
resistivity has to be measured directly with and without the presence of SB.
We have therefore devised the following experiment. Four electrical contacts
were deposited in the central part of large square-shaped BSCCO crystals, as
shown schematically in the inset to Fig. 1, and transport measurements were
carried out in this `unconventional' geometry. Then the crystals were cut
along the marked `cut lines' in Fig. 1 to form the `standard' narrow strip
geometry, and remeasured.

The described findings were confirmed on several high quality 
BSCCO single crystals ($T_c\simeq $89 K).  Here we present results 
for two of the crystals, grown by the traveling solvent floating 
zone method \cite{motohira}. Crystal A was cleaved and cut to 
dimensions of 1.2 mm ({\it w})$\times 1.1 $ mm ({\it l}) in the {\it 
ab} plane and $13\mu {\rm m}$ thick, and was subsequently cut to a 
0.36 mm wide strip. Crystal B was initially $1.3$ ({\it w})$\times 
1.7$ ({\it l}) $\times 0.013$ ({\it d}) mm$^{3}$, and was later cut 
to 0.29 mm width. Four Ag/Au pads for electrical contacts were 
thermally evaporated and had dimensions of $100\times 200\mu m^{2}$ 
with $75\mu m$ separation.  The same contacts were used for 
transport measurements in the `square' and `strip' geometries. The 
magnetic field was applied parallel to the crystalline {\it c}-axis, 
and the four-probe resistance was measured using an ac bridge.

Cutting the crystals is expected to increase their resistance because the
cross section is reduced. This is indeed the case at temperatures above 
$T_{c}$ where the resistance of the strip crystals is typically a factor of
three larger than before cutting, consistent with the change in the
geometry. In contrast, at lower temperatures the cutting {\em decreases} the
measured resistance dramatically, by as much as two orders of magnitude, as
shown in Fig. 1 for various fields and temperatures. If the sample
resistance is governed by bulk properties, any reduction of the cross
section must necessarily lead to an increase of the measured resistance, in
striking contrast to the experimental result. The observed large decrease of
the resistance of the strip crystals is a direct manifestation of the
dominant role of the SB.

In the standard strip geometry in perpendicular field the flowing vortices
enter the sample from one edge, flow across the bulk, and exit at the other
edge. This process involves vortex activation over the SB and drift of
vortices across the bulk. Electrically this situation can be represented by
parallel resistors, $R_{S}$ of the edges and $R_{B}$ of the bulk \cite
{burlachkov}. A large SB impedes vortex transmission through the edges and
is equivalent to a low $R_{S}$, which shunts the relatively large bulk 
$R_{B} $. As a result, the measured sample resistance $R_{S}||R_{B}$ is
reduced, and most of the current flows along the low resistance 
edges of the sample as was demonstrated by the self-field 
measurements \cite{sf,pd}.  In order to measure the true bulk 
transport properties one has to use a configuration in which 
vortices circulate in the bulk without crossing the edges. This 
situation is approached here by simply using the large square 
crystals. Since the current contacts are positioned far from all the 
edges, most of the vortices circulate around the two current 
contacts, and the measured voltage drop between the voltage contacts 
is thus determined by bulk vortex properties. Even if the SB is 
extremely large, so that $R_{S}$ vanishes along the entire 
circumference, the measured resistance is reduced only slightly as 
compared to an infinite sample with no edges. This is because the 
distance between the two current contacts is comparable or smaller 
than their distance from the edges. Hence any current path that 
includes a segment along the low resistance edges also includes the 
high resistance segments from the injection contact to the edge and 
back from the edge to the drain contact. In the strip crystals, on 
the other hand, the contacts are in close proximity to the edges and 
thus most of the current is readily shunted by the low resistance 
edges. Our numerical calculations of current distributions
confirm this description and show 
that the measured resistance of the square crystal reflects mainly 
the bulk resistance, whereas the resistance in the strip geometry is 
dominated by the SB resistance\cite{Funp}.  
The SB can also be avoided in the Corbino disk geometry. 
Measurements showing the same effects as in Figs. 1 to 5 will be 
presented elsewhere \cite{Dunp}.

We now analyze the results in more detail. In Fig. 1 the resistance values
have been normalized to the normal state resistance for clarity. At $T=95$K
the resistances are 0.23 and 0.62 $\Omega $ before and after cutting,
respectively. Below $T_{c}$, however, the effect of the SB on reducing the
resistance of the strip crystal is very pronounced. The maximum relative
change in the resistance is observed at fields of the order of 500 Oe,
decreases gradually at higher fields, and tends to become negligible at
fields of about 2T. Such a decrease of the SB with increasing field is
expected theoretically \cite{Clem}. Another important observation is that
the relative change in the resistance generally grows with decreasing
temperature. At low temperatures the resistance is commonly expected to be
governed by bulk pinning, which should dominate over the SB. However, this
is just contrary to the behavior observed in Fig. 1: Thermal activation over
the SB, which decreases with decreasing temperature, remains the
dominant limitation for vortex motion in the strip samples, and even becomes
more important, as the temperature is decreased. In other words,
whenever vortex flow is large enough for a finite resistance to be detected,
the corresponding measured resistance in the strip geometry is 
governed by the SB rather than the bulk properties, except close to 
$T_{c}$ or at high fields.

In addition to altering the value of the resistance, the SB also governs the
measured activation energy. Figure 2 shows the effective activation energy 
$U$
as obtained by a linear fit to the Arrhenius data of Fig. 1 (for fields
below 1000 Oe the Arrhenius part in the vortex-liquid phase above the
melting kink is used). The activation energy of the strip is significantly
larger than that of the square crystal indicating that $U$ of the SB is
significantly larger than the vortex pinning $U$ in the bulk. Since in both
geometries the bulk and surface contributions are partially intermixed, we
believe that the strip values of $U$ reflect a lower bound for the SB
activation energy, whereas the square sample $U$ is the upper bound for the
bulk activation. The activation energies of our strip samples are comparable
to the values reported in the literature \cite{palstra,busch}.

The SB also dominates the nonlinear characteristics of the resistive
behavior. Figure 3 shows the normalized $R(T)$ of crystal B at 700 Oe for
various currents. The square crystal, governed by the bulk properties,
displays nearly linear resistance, except at the lowest temperature, as
shown for the 5 and 15 mA current (thick curves). The SB dominated strip
crystal, in contrast, shows highly nonlinear resistance at low currents, as
seen in Fig. 3, which tends to approach the bulk value as the current is
increased. At low currents the transmissivity of the SB is low and thus the
resistance of the strip crystal is significantly reduced. The height of the
SB decreases with increasing transport current, resulting in a nonlinear
resistance \cite{burlachkov}. This fundamental property of the SB is the
main source of the apparent nonlinear resistive behavior in the
vortex-liquid phase, which was previously ascribed to a liquid state with
high bulk viscosity \cite{tsuboi}.

We now focus on the resistive kink at the first-order vortex lattice phase
transition (FOT) \cite{safar,kwok}. Upon freezing, the resistance in 
BSCCO crystals is observed to drop sharply below the noise level 
\cite {simultaneous-watauchi,sublimation}. This behavior is often 
ascribed to the sudden onset of bulk pinning. We now proceed to 
evaluate the contribution of the SB to this resistive transition. 
Figure 4(a) shows $R(T)$ of crystal B at 300 Oe in the vicinity of 
the FOT at a relatively high current of 10mA in order to resolve the 
behavior in the solid phase. The square crystal reflects the 
behavior of the bulk resistance $R_{B}$ which shows an appreciable 
drop upon freezing. However, the important observation here is that 
the strip sample exhibits an even larger drop and a significantly 
lower resistance in the solid phase. This means that in the strip 
samples the SB dominates the resistive behavior also in the solid 
phase. Figure 4(b) shows the ratio of the normalized resistances 
$R_{square}/R_{strip}$ which reflects the ratio $R_B/R_S$. In the 
fluid the edge resistance $R_S$ is lower than $R_B$ resulting in a 
ratio $R_{square}/R_{strip} > 1$ . However, upon freezing, $R_S$ 
drops dramatically resulting in as much as two orders of magnitude 
increase in $R_{square}/R_{strip}$ in Fig 4(b). This result 
demonstrates that the SB itself undergoes a sharp transition from a 
relatively low barrier height in the fluid phase to significantly 
enhanced barrier in the solid phase. Such a transition is indeed 
expected theoretically for the melting (vortex solid to line liquid) 
and, even more so, for the sublimation (vortex solid to pancake gas) 
transition \cite{burlachkov,koshelev}. In the pancake gas the vortex 
penetration is accomplished by activation of individual pancakes, which is a
relatively easy process. In the solid phase, on the other hand, an extended
line vortex has to penetrate and to distort the solid lattice in the bulk,
which results in large activation energy \cite{koshelev}. Figure 4 is
therefore a direct manifestation of the sharp transition of the SB that
occurs concurrently with the bulk FOT. This finding is consistent with the
conclusions from the self-field studies \cite{pd}. Upon freezing both $R_B$
and $R_S$ drop sharply. However, since $R_S$ dominates the resistance both
above and below the FOT, it is important to realize that the resistive
transition that is usually measured in strip-shaped BSCCO crystals mainly
reflects the sharp drop of $R_S$, rather than of $R_B$ which is obscured by
the SB.

The particularly large SB in the vortex-solid phase results usually in
immeasurably low resistance below the FOT. The use of the square crystals
provides therefore a unique method to study the true bulk vortex properties
in the quasi-lattice or Bragg glass phase 
\cite{nattermann,gl,kierfeld-knh,ertas-vinokur} below the FOT.  
Figure 5 shows an example of the resistive behavior of square 
crystal at 100 Oe for various currents. In contrast to the strip 
crystal for which the apparent resistance drops rapidly to below the 
noise level (dashed curve), the bulk resistance of the square sample 
can be measured down to as much as 20 K below the FOT at elevated 
current.  In addition, in contrast to the vortex-fluid phase, which 
shows almost linear behavior above the FOT, the quasi-lattice 
displays a highly nonlinear resistivity. The bulk resistance is 
thermally activated with activation energy being strongly current 
dependent.  The inset to Fig. 5 shows the values of $U$ obtained by 
a linear fit to the Arrhenius data for 100, 300, and 500 Oe. The 
activation energy has approximately power-law dependence, $U \propto 
I^{-\mu}$, with $\mu \approx 0.5$.
This power-law current dependence in the vortex-solid is expected 
from recent theoretical treatments of the Bragg glass phase 
\cite{review,nattermann,gl,kierfeld-knh,ertas-vinokur}. 
Furthermore, the same theories predict a weak  field dependence of 
$U$, also in agreement with the inset to Fig. 5. 
These results, without the influence of 
the SB, provide therefore a first experimental test of 
the Bragg glass theories by transport measurements in BSCCO. 

In summary, we have demonstrated that in the conventional strip geometry the
SB dominates the resistive behavior of BSCCO crystals over a wide range of
temperatures and fields. The SB governs the value of the resistance, its
nonlinear behavior, the apparent activation energy, and the resistance drop
at the first-order transition. The true bulk properties can be probed using
an alternative configuration in which the electrical contacts are well
separated from the edges of large crystals. This geometry provides an
important new access for investigation of the transport properties in the
Bragg glass phase below the FOT.

Helpful discussions with V. B. Geshkenbein, T. Giamarchi, P. Le 
Doussal, T. Nattermann and J. Kierfeld are gratefully acknowledged. 
This work was supported by the Israel Ministry of Science and the 
Grant-in-Aid for Scientific Research from the Ministry of Education, 
Science, Sports and Culture, Japan, by the Israel Science 
Foundation, and by the MINERVA Foundation, Munich, Germany.


\newpage 

\ FIGURE CAPTIONS

Fig. 1. Arrhenius plot of the normalized resistance of BSCCO crystal A in
the square geometry (thick curves), and after cutting into a strip (thin
curves), for various applied fields between 92 and 5000 Oe ($I=1$ mA).
Inset: Schematic sample geometry. Four probe resistance was measured on a
square crystal (thick perimeter) with contacts deposited in the center. The
crystal was then cut into a strip along the marked lines (thin lines) and
remeasured.

Fig. 2. The activation energy $U$ as a function of applied field for the
square ($\Box$) and strip ($\circ$) sample geometries. The activation
energies were obtained by a linear fit to the Arrhenius part in the
vortex-liquid phase of Fig. 1.

Fig. 3. Arrhenius plot of the normalized resistance of crystal B at 700 Oe.
Thick curves: resistance of the square sample at 5 and 15 mA current. Thin
curves: nonlinear resistance in the cut strip geometry for 0.5 to 15 mA
currents.

Fig. 4. (a) Normalized $R(T)$ of the square and strip crystal B in the
vicinity of the FOT at 300 Oe and 10 mA current. (b) The ratio of the
normalized square and strip resistances from (a).

Fig. 5. Arrhenius plot of the normalized resistance of the square sample B
at 100 Oe and various currents between 1 and 30 mA. 
The dashed curve is the normalized
resistance of the strip at 3 mA for comparison. Inset: Activation energy as
a function of current in the vortex-solid phase at applied fields of 100 Oe
($\bullet$), 300 Oe ($\times$), and 500 Oe ($\Box$). Solid line: 
$U = C I^{-0.5}$ with $C$=5200 K mA$^{0.5}$.

\end{multicols}

\end{document}